\documentclass[reprint,twocolumn,prl,aps,showpacs,letterpaper,superscriptaddress]{revtex4-1}
\usepackage[T1]{fontenc}
\usepackage[applemac]{inputenc}
\usepackage{verbatim}
\usepackage{prettyref}
\usepackage{amstext}
\usepackage{amssymb}
\usepackage{graphicx}
\usepackage{ulem}
\usepackage{hyperref}
 \hypersetup{colorlinks=true,citecolor=blue}

\usepackage[nice]{nicefrac}
\usepackage{subfig}

\usepackage[usenames,dvipsnames]{color}
\makeatletter

\@ifundefined{textcolor}{}
{
 \definecolor{BLACK}{gray}{0}
 \definecolor{WHITE}{gray}{1}
 \definecolor{RED}{rgb}{1,0,0}
 \definecolor{GREEN}{rgb}{0,1,0}
 \definecolor{BLUE}{rgb}{0,0,1}
 \definecolor{CYAN}{cmyk}{1,0,0,0}
 \definecolor{MAGENTA}{cmyk}{0,1,0,0}
 \definecolor{YELLOW}{cmyk}{0,0,1,0}
 }

\makeatother
\usepackage[english]{babel}

\begin{document}
\selectlanguage{english}

\title{Entanglement irreversibility from quantum discord and quantum deficit}

\author{Marcio F. Cornelio}
\email{mfc@ifi.unicamp.br}
\affiliation{Instituto de F\'isica Gleb Wataghin, Universidade Estadual de Campinas, CEP 13083-859, Campinas, S\~ao Paulo, Brazil}

\author{Marcos C. de Oliveira}
\email{marcos@ifi.unicamp.br}
\affiliation{Instituto de F\'isica Gleb Wataghin, Universidade Estadual de Campinas, CEP 13083-859, Campinas, S\~ao Paulo, Brazil}

\author{Felipe F. Fanchini}
\affiliation{Instituto de F\'isica Gleb Wataghin, Universidade Estadual de Campinas, CEP 13083-859, Campinas, S\~ao Paulo, Brazil}
\affiliation{Departamento de F\'isica, Universidade Federal de Ouro Preto, CEP 35400-000, Ouro Preto, MG,
Brazil}

\begin{abstract}
We relate the problem of irreversibility of entanglement with the recently
defined measures of quantum correlation - quantum discord and one-way
quantum deficit. We show that the entanglement
of formation is always strictly larger than the coherent information
and the entanglement cost is also larger in most cases. We
prove irreversibility of entanglement under LOCC for a family
of entangled states. This family is a generalization of the maximally
correlated states for which we also give an analytic expression for the distillable entanglement, the relative entropy of entanglement, the distillable secret key and the quantum discord.
\end{abstract}

\maketitle

Two complementary and among the most important tasks in quantum
information theory (QIT) are entanglement dilution and entanglement
distillation \cite{Bennett-etal96-pra,BennettEtal96-PhysRevA.54.3824}.
These tasks are performed in a scenario where two spatially separated observers, usually called
Alice and Bob, share some quantum states and are able to manipulate
their respective parties through local operations and classical communication (LOCC) \cite{BennettEtal96-PhysRevA.54.3824}. 
In the first task, Alice and Bob share a large number of copies of a standard pure maximally entangled state,
\begin{equation}
|\Phi\rangle=\frac{1}{\sqrt{2}}(|00\rangle+|11\rangle),
\label{ebit}
\end{equation}
which is associated with a unit of entanglement called \textit{e-bit}. Their
task is to construct many copies of an arbitrary, generally mixed, state $\rho$ from many copies of  $|\Phi\rangle$ 
using only LOCC (See Fig. 1). In the second task, Alice and Bob want to perform
the reverse operation, i. e., to extract from many copies of an arbitrary
state, generally mixed, the maximal possible amount of e-bits using
only LOCC.
Those tasks naturally raise the two most important
measures of entanglement - \textit{entanglement cost}
($E^{\cal C}$) and  \textit{distillable entanglement} ($E^{\cal D}$)
\cite{BennettEtal96-PhysRevA.54.3824}. For a given state $\rho_{ab}$,
$E^{\cal C}(\rho_{ab})$ is the optimal rate for converting a large number
of e-bits into a large number of copies of the mixed state $\rho_{ab}$ under LOCC by Alice and Bob. Similarly $E^{\cal D}(\rho_{ab})$ is the
optimal rate for converting a large number of $\rho_{ab}$ into e-bits
under LOCC \cite{HorodeckiReview09}.

When Alice and Bob can build a large number of copies of an arbitrary
state $\rho_{ab}$ and can get the same amount of e-bits back through LOCC,
it is said that there is  \textit{entanglement reversibility}.
Conversely, the entanglement is said \textit{irreversible}.
To understand the aspects leading to entanglement irreversibility is one of the most important open problems in QIT \cite{BennettEtal96-PhysRevA.54.3824} with practical implications.
Particularly, entanglement dilution is connected to the problem of classical communication over a noise quantum channel \cite{Shor04-additivy,*CornelioOliveira10} and entanglement distillation is connected to quantum communication and quantum key distribution \cite{HorodeckiReview09,DevetakWinter04,*DevetakWinter05,Horodecki00-noisycoding,DevetakWinter04-ccr}  for secure cryptography.
\begin{figure}
      \includegraphics[width=0.48\textwidth]{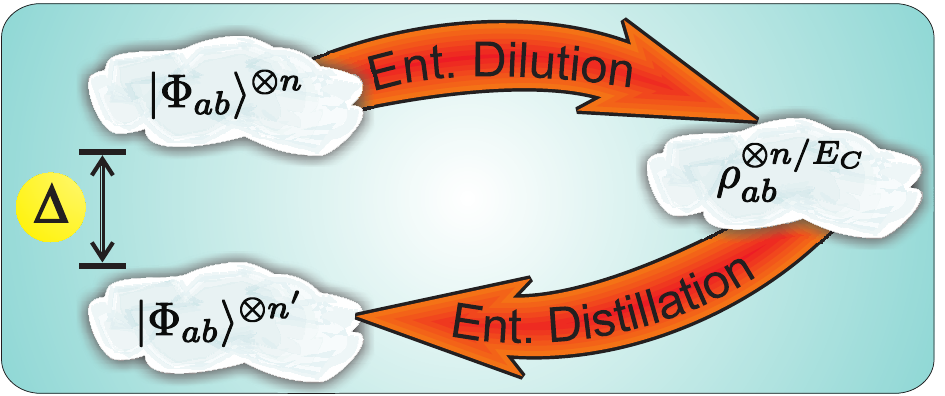}
       Figure 1: (Color on-line) Entanglement Dilution-Distillation cycle. The entanglement loss is given by $\Delta$. In the case of reversible entanglement, $\Delta$ vanishes. In the irreversible case of Eqs. (\ref{eq:FundAsymp},\ref{eq:EcDefEd}), $\Delta$ is  the regularized quantum discord.
\end{figure}    
It is known that the task of building an entangled state and extracting back the e-bits is reversible if Alice and Bob are limited to build and to distill pure entangled
states  \cite{Bennett-etal96-pra}.
For a pure state $\varphi$, $E^{\cal C}$ and $E^{\cal D}$ are equal to the von Neumann entropy $S(\rho_r)$ of the reduced density matrix $\rho_r$ of one of the subsystems.
Moreover it is {a long-standing conjecture}
that the only states with $E^{\cal C}=E^{\cal D}$ are pure
states and the so-called pseudo-pure (PP)  \cite{Horodecki98,HorodeckiReview09} states, \begin{equation}
\rho_{pp}=\sum p_{i}|\varphi_{ab}^{i}\rangle\langle\varphi_{ab}^{i}|\otimes|f_{i}\rangle\langle f_{i}|\label{eq:pseudoPuro},\end{equation}
where $|f_{i}\rangle$ is an \textit{ancilla}, 
locally accessible for Alice or Bob, working as a flag that indicates which pure entangled state $|\varphi_{ab}^{i}\rangle$ is in the mixture. 
Although widely believed, there are few concrete evidences for this conjecture. 
To understand irreversibility for mixed states has revealed itself a very difficult question and the first examples were given 
much later in Refs.  \cite{VidalCirac01-prl,*VidalCirac01-PRA,Vidal02-prl,VollbrechtWerner04-PhysRevA.69.062304,Yang-Horodecki05}.
Particularly, in Ref. \cite{Yang-Horodecki05} it is shown that one can find mixed states that consume entanglement to be created but no entanglement can be extracted from it, the so-called bound entanglement.

One of the main reasons why it is so difficult to understand irreversibility
for mixed states is that $E^{\cal C}$ and $E^{\cal D}$ are given by formal
limits that are very hard to evaluate in general.
The first attempt to quantify the entanglement cost was given by Bennett \textit{et al}  \cite{BennettEtal96-PhysRevA.54.3824} introducing the \textit{entanglement of formation} (EOF),  
\[
E^{\cal F}(\rho)=\min_{\mathcal{E}}\Big\{\sum_{i}p_{i}E^{\cal C}(\varphi_{i})\Big\},
\]
where the minimization is over the set $\mathcal{E}$ of all ensembles of pure states $\{p_i,\varphi_i\}$ such that $\rho=\sum_{i}p_{i}\varphi_{i}$.
EOF is the cost of diluting the e-bits in the pure states of the ensemble of $\rho$  and mixing them. As there are many ensembles that realizes  $\rho$, one can always choose the ensemble that gives the minimal cost, hence the minimization in the formula of $E^{\cal F}$. 
For a long time, it was generally believed that this method was the best dilution protocol and that $E^{\cal C}=E^{\cal F}$. {Indeed}, it was shown by Hayden  \textit{et al.} \cite{Hayden01-JPhysA34} that 
 $E^{\cal C}$ is the  \textit{regularization} of the EOF: 
\begin{equation}
E^{\cal C}(\rho)=\lim_{n\rightarrow\infty}\frac{1}{n}E^{\cal F}(\rho^{\otimes n}).
\label{Ecost}
\end{equation}
So the question was reduced to whether $E^{\cal F}(\rho^{\otimes n})=nE^{\cal F}(\rho)$ or not, that is, whether EOF is an additive measure
\cite{Shor04-additivy,*CornelioOliveira10}.
However, recently it was shown that EOF is not additive in general \cite{Hastings2009}, implying that there are states for which better dilution protocols exist than the one given the EOF.
For such states, $E^{\cal F}(\rho^{\otimes n})<nE^{\cal F}(\rho)$ for some $n$ and $E^{\cal C}$ is strictly smaller than $E^{\cal F}$. 
Since $E^{\cal F}$ is known to be additive only for very particular states \cite{Vidal02-prl,DevetakWinter04-ccr},
 it is not generally  known when one can take $E^{\cal F}$ for $E^{\cal C}$.

The difficulty is similar for evaluating $E^\mathcal{D}$. In fact, $E^{\cal D}$ is only known in the particular case of maximally correlated (MC) states \cite{Rains09-PhysRevA.60.179}.
There is an important lower bound, however.
When one of the conditional entropies $S_{a|b}$ or $S_{b|a}$ is negative, ($S_{a|b}=S_{ab}-S_b$), there is a protocol called \textit{hashing} which can distill $-S_{a|b}$ e-bits from $\rho$ \cite{BennettEtal96-PhysRevA.54.3824,DevetakWinter04,*DevetakWinter05}. Then the  \textit{coherent information},  $I_{C}=\max\{0,-S_{a|b},-S_{b|a}\}$ capture this negative part and is a lower bound for $E^{\cal D}$.
Indeed it is known that $I_{C}$ can be increased by LOCC and notably an optimal 
distillation protocol can always be achieved  performing the optimization of $I_{C}$ followed by hashing \cite{DevetakWinter04,*DevetakWinter05}. That is
\begin{equation}
E^{\cal D}(\rho)=\lim_{n\rightarrow\infty}\sup_{V}\frac{1}{k}I_{C}\left(V\rho^{\otimes k}\right),
\label{eq:Ed}
\end{equation}
where $V$ is some LOCC operating on $k$ copies of $\rho$.
There is no bound on the number of copies $V$ can act.
So  $E^{\cal D}$ might in fact exist only as the limit of $V$ acting on a very large number of copies of $\rho$. In the end, it is very difficult to know or to efficiently bound $E^{\cal C}$ and $E^{\cal D}$ simultaneously for answering the reversibility question. The difficulty in calculating these quantities is the main reason for this questioning to be open for 14 years \cite{BennettEtal96-PhysRevA.54.3824}.
Here we will be able to calculate $E^{\cal D}$ for a new family of states.

In this context, it is convenient to introduce our first formal results in the form of an important Theorem and a Lemma.
In what follows, when we say a mixed state, we mean a \textit{not pure and not PP} state.

\textit{Lemma 1}: For every mixed entangled state $\rho_{ab}$
\[
E^{\cal F}(\rho_{ab})>I_{C}(\rho_{ab}),
\]
i.e., the EOF of is strictly larger than the coherent information for every mixed $\rho_{ab}$.

\textit{Theorem 1}: 
Let $\rho_{ab}$ be a mixed entangled state, if 
\begin{eqnarray}
E^{\cal C}(\rho_{ab}) & = & \frac{1}{n}E^{\cal F}(\rho_{ab}^{\otimes n})
\label{CondA}
\\
E^{\cal D}(\rho_{ab}) & = & \max_{V}\frac{1}{k}I_{C}\left(V\rho^{\otimes k}\right)
\label{CondB}
\end{eqnarray}
for some finite $n$ and $k$, then the entanglement is irreversible for $\rho_{ab}$, i. e., $E^{\cal C}(\rho_{ab})>E^{\cal D}(\rho_{ab})$.

The technical details of the proofs of Lemma 1 and Theorem 1 are left to the supplementary material 
 Here we limit ourselves to discuss their meaning in the context of entanglement irreversibility and the main concepts involved.
First we notice that Eq. (\ref{CondA}) and Eq. (\ref{CondB}) differ from Eq. (\ref{Ecost}) and Eq. (\ref{eq:Ed}) only by the lacking of the limits.
So entangled states satisfying condition (\ref{CondA}) will be called \textit{type} $A$  and satisfying condition  (\ref{CondB}) will be called \textit{type} $B$.
 The states satisfying both conditions will be called \textit{type} $AB$ and, to complete the analogy, states satisfying none will be called \textit{type}  $O$.
 In this way, the Theorem 1 simply says that states of type $AB$ are irreversible.
It is important to notice that for all states  that $E^{\cal C}$ and/or $E^{\cal D}$ are known, the conditions  (\ref{CondA}) and/or  (\ref{CondB}) are satisfied.

The central concept behind Lemma 1 and Theorem 1 is the quantum discord \cite{zurek01,*hendersonvedral01}.
It is defined as the difference between two ways of defining mutual information,
\[
\delta_{a|c}(\rho_{ac})=I(\rho_{ac})-J_{a|c}(\rho_{ac}),
\] 
where $I(\rho_{ac})=S(\rho_a)+S(\rho_c)-S(\rho_{ac})$  is the quantum mutual information and $J_{a|c}(\rho_{ac})$ is a measure of the amount of classical correlations present in quantum states,
\[
 J_{a|c}(\rho_{ac})=\max_{\{ \Pi_i\} }\big{[}S(\rho_{a})-\sum_{i}p_{i}S(\rho_{a}^{i}|\Pi_{i})\big],
\]
where $\{\Pi_{i}\}$ is a complete POVM on subsystem $c$ and $p_{i}$ are the respective probabilities, so $S(\rho_{a}^i|\Pi_{i})$ is the entropy of subsystem $a$ conditioned to the output $\Pi_{i}$ on $c$. So $I(\rho_{ac})$ measure the total amount of correlations in $\rho_{ac}$ while $J_{a|c}(\rho_{ac})$ measures the amount of classical correlations when the POVM $\{\Pi_{i}\}$ is performed on $c$. In this way, $\delta_{a|c}(\rho_{ac})$ gives a distinct notion of non-classicality from entanglement.

It is easy to relate quantum discord with the EOF. For every pure tripartite state $|\psi_{abc}\rangle$ holds \cite{koashiwinter04-physreva.69.022309,*FanchiniCornelio10}
\begin{equation}
E^{\cal F}(\rho_{ab})  =  \delta_{a|c}(\rho_{ac})-S_{a|b}(\rho_{ab})
\label{eq:FundNonAsymp}
\end{equation}
where $\rho_{ab}$ and $\rho_{ac}$ are the reduced states of the respective subsystems. From Eq. (\ref{eq:FundNonAsymp}) it is easy to see that $ \delta_{a|c}$ is not additive only when $E^{\cal F}(\rho_{ab})$ is not additive as well. Then it is necessary to define the \textit{regularizated quantum discord} (RQD) in the same way as for $E^{\cal F}$,
\[
\Delta_{a|c}(\rho_{ac})=\lim_{n\rightarrow\infty} \frac{1}{n} \delta_{a|c}(\rho_{ac}).
\]
Similarly  to Eq. (\ref{eq:FundNonAsymp}), we have for the regularized quantities
\begin{equation}
E^{\cal C}(\rho_{ab})  = \Delta_{a|c}(\rho_{ac})-S_{a|b}(\rho_{ab}),
\label{eq:FundAsymp}
\end{equation}

Eq. (\ref{eq:FundAsymp}) is relating three fundamental quantities in QIT with a clear operational meaning.
It is known that when the conditional entropy is negative it is possible to distill $-S_{a|b}$ e-bits out of  the state $\rho_{ab}$.
Then Eq. (\ref{eq:FundAsymp}) is telling us that the amount of entanglement lost in the process of creating a mixed state $\rho_{ab}$ and distill it by hashing is equivalent to the RQD with a complementary system $c$.
Thus Eq. (\ref{eq:FundAsymp}) gives a  new operational meaning to $\Delta_{a|c}$ as \textit{ a measure of the amount of entanglement loss when Alice and Bob distill entanglement by hashing}.

For states of type $B$ {and $AB$}, i.e., all those satisfying condition (\ref{CondB}),  the connection between the RQD with the purifying subsystem $c$ and entanglement loss in distillation will turn clear.  For every $\rho_{ab}$ of type $B$ there is a finite $k$ and a LOCC $V^\prime$ giving the maximum in Eq. (\ref{CondB}) such that
\begin{equation}
\Delta_{a|c}(\sigma_{ac})=E^{\cal C}(\sigma_{ab})-E^\mathcal{D}(\sigma_{ab}),
\label{eq:EcDefEd}
\end{equation}
where $\sigma_{ab}=V^\prime \rho_{ab}^{\otimes k}$ and $E^{\cal D}(\sigma_{ab})=kE^{\cal D}(\rho_{ab})$.  In this way, we say that $\sigma_{ab}$ is the optimized distillable state (ODS) of $\rho_{ab}$. We notice that $\sigma_{ab}$ can be the ODS of many distinct states, being the result of also distinct $V^\prime$s. Therefore each  $\sigma_{ab}$ satisfying Eq. (\ref{eq:EcDefEd}) defines a class of states $\rho_{ab}$ for which it is the ODS.
For each class, $\Delta_{a|c}(\sigma_{ac})$ is the minimal amount of entanglement lost in any distillation protocol for all states belonging to the class.
In the case of $\rho_{ab}$ being bound entangled, we have for any  $\sigma_{ab}=V\rho_{ab}$, with an arbitrary LOCC $V$, that
\[
\Delta_{a|c}( \sigma_{ab}) \ge E^{\cal C}(\rho_{ab}).
\]

We have stated our more general results. Now we apply these results for an important case - We consider the tripartite state
\[
|\psi_{abc}\rangle = \sum_{i=1}^N \alpha_i|a_i, i_b, c_i \rangle
\] 
where $N$ is the dimension of the subsystems, $\{|i_b\rangle\}$ is an orthonormal basis for $b$, $\{|a_i\rangle\}$ and $\{|c_i\rangle\}$ are arbitrary (usually non-orthogonal) states of $a$ and $c$. The subsystem $ab$ results in the density matrix 
\begin{equation}
\rho_{ab}=\sum_{ij}\beta_{ij}|a_{i}i_b\rangle\langle a_{j}j_b|,
\label{1-MC}
\end{equation}
where $\beta_{ij}=\alpha_i\alpha_j^*\langle c_j|c_i\rangle$. We call these states \textit{one-way maximally correlated} (1-MC) since, despite $\rho_{ab}$ being mixed, the result of a measurement in the basis $\{|i_b\rangle\}$ is perfectly correlated with a definite state $|a_i\rangle$.

\textit{Theorem 2}: For every mixed 1-MC $\rho_{ab}$ the entanglement is irreversible. Eq. (\ref{eq:EcDefEd}) holds and
\[
E_{ab}^{\cal C}>E^\mathcal{D}_{ab}=\delta_{a|b}=\Delta_{a|b}=-S_{a|b}.
\]

In fact, 1-MC states are examples of type $AB$ states.
The essential elements of the proof are the fact that EOF is additive for them and the distillable entanglement turns out to be exactly $-S_{a|b}$. Furthermore, Theorem 2 gives us also the quantum discord in one direction, $\delta_{a|b}$ (as well as other measures, {like} the relative entropy of entanglement and the distillable secret key, see the supplementary text for details) for these states. From the fact that  $\delta_{a|b}\geq 0$  one can deduce that $-S_{a|b}\geq0$.
We know also that $\delta_{a|b}=0$ implies that $\rho_{ab}$ is separable.
So \textit{$-S_{a|b}=0$ is a necessary and sufficient separability criteria for 1-MC states} and there is no bound entangled state belonging to this family.

In addition, we should notice that the only examples of irreversibility previously known \cite{VollbrechtWerner04-PhysRevA.69.062304} with $E^{\cal D}>0$ are very particular cases of 1-MC states.
Furthermore also the examples for which we knew $E^{\cal D}$ \cite{DevetakWinter04,*DevetakWinter05,Rains09-PhysRevA.60.179} are also a subset of null measure of 1-MC states. Therefore, the only states proved irreversible are the bound entangled and 1-MC correlated  states.
\begin{figure}
\subfloat[$\varphi=\nicefrac{\pi}{6}$]{\includegraphics[scale=0.37]{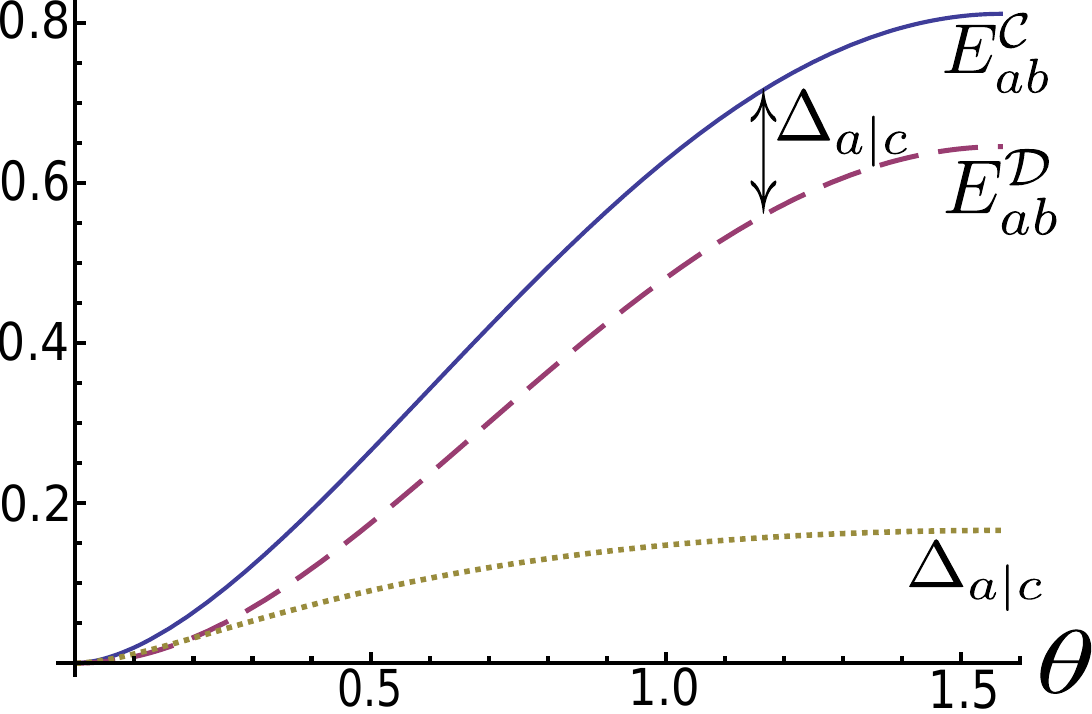}

}\subfloat[$\varphi=\nicefrac{\pi}{4}$]{\includegraphics[scale=0.37]{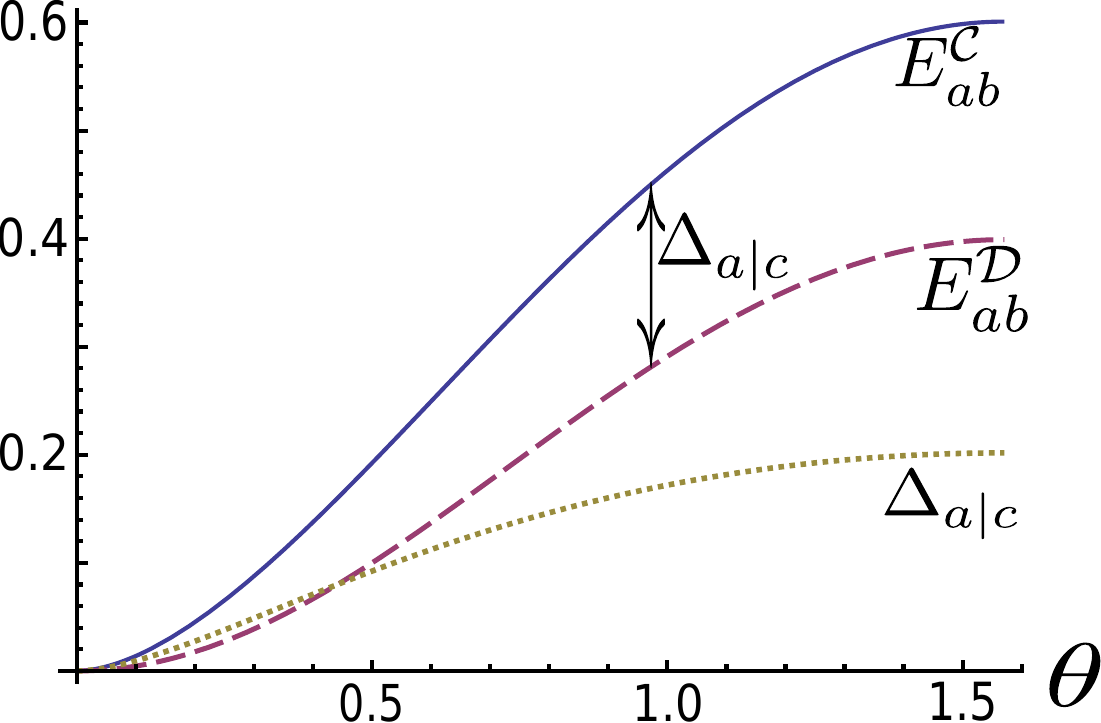}

}

Figure 2: (Color on-line) $E^{\cal C}$, $E^{\cal D}$ and $\Delta_{a|c}$ for 1-MC states of two-qubits $\sigma_{ab}$. The values satisfy Eq. (\ref{eq:EcDefEd}) and $E^{\cal D}$ was previously known only for the case $\theta=\nicefrac{\pi}{2}$.

\end{figure}

\textit{Example} - A tripartite pure states satisfying the condition of the Theorem 2, i. e., such that the reduced state $\rho_{ac}$ is separable, can be written as
\[
|\psi_{abc}\rangle = \frac{1}{\sqrt2}(|000\rangle + |\theta1\varphi\rangle)
\]
where $|\theta\rangle=\cos\theta |0\rangle+\sin\theta|1\rangle$ and $|\varphi\rangle=\cos\varphi |0\rangle+\sin\varphi|1\rangle$
The resulting 1-MC is given by
\[
\sigma_{ab}=\frac{1}{2}\big[
|00\rangle\langle00| +|\theta 1\rangle\langle \theta 1|
 + \cos\varphi(|00\rangle\langle \theta 1| +|\theta 1\rangle\langle00|)
\big].
\] 
Notice that a measurement in the basis $\{|0\rangle,|1\rangle\}$ has a perfect correlation with the states $|0\rangle$ and $|\theta\rangle$.
The angle $\theta$ gives how far is $\sigma_{ab}$ from an usual  MC state \cite{Rains09-PhysRevA.60.179} belonging to this class when $|\theta\rangle=|1\rangle$.
The angle $\varphi$ gives the amount of mixedness of $\rho_{ab}$. For $\varphi = 0$ the state is pure and for $\varphi = \nicefrac{\pi}{2}$ the state is separable. Figure 2 shows the behavior of $E^{\cal C}_{ab}$ and $E^{\cal D}_{ab}$ and how the loss of entanglement is equal to $\Delta_{a|c}$. \textcolor{black}{It is remarkable that this class is now the only one for which we know both $E^{\cal C}$ and $E^{\cal D}$.}

Combining Theorems 1 and 2 we easily get:

\textit{Corollary 1}: A type $B$ reversible mixed state $\rho_{ab}$ exists \textit{if and only if} there exists a bound entangled state $\rho_{ac}$ such that $\delta_{a|c}>0$ and $\Delta_{a|c}=0$.

The question about existence of states with  $\Delta_{a|c}=0$, but $\delta_{a|c}>0$, has already been raised in 2005 \cite{horodeckietal05-deficit-pra} and 
 is directly related to question of additivity of EOF by Eq. (\ref{eq:FundNonAsymp}). So our results tell us exactly in which situation the non-additivity of EOF could be responsible for irreversibility providing a strong link between these two fundamentals questions.

\textit{Thermodynamical analogy} - 
Since the beginning of entanglement theory, it has been compared with thermodynamics.
Dilution and distillation of pure into pure states are reversible operation under LOCC.
This is analogous to a reversible process in classical thermodynamics where entropy remains constant and all the energy that is put in the system can be recovered without losses.
Mixedness is caused by some noise and is associated with the increasing of entropy. Then our intuition tell us that noise probably implies in some irreversible loss of entanglement that cannot be recovered by LOCC only. However this connection has never been done explicitly.
Our work provides the desired connection directly between that noise and entanglement loss.

Zurek  \cite{zurek2003} has shown that QD can interpreted as some amount of thermodynamical work that Alice and Bob must pay when they operate only locally on their respective subsystems.
The same operational interpretation was developed independently \cite{oppenheimhorodecki02-workdeficit,*devetak05-pra,horodeckietal05-deficit-pra} generating many kinds of a similar quantity called \textit{quantum deficit}. In the asymptotic limit, the regularized expressions for QD and one-way quantum deficit are equivalent.
The quantum deficit measures the following task: Suppose that Alice and Bob share many copies of $\rho_{ab}$. From that they can use the information they have about this state to produce work through a Szilard engine  \cite{oppenheimhorodecki02-workdeficit,*devetak05-pra,horodeckietal05-deficit-pra}.
However there is a difference between the amount of work Alice and Bob can perform whether they operate globally with the two subsystem or they can operate only locally on its respective subsystems. This difference in the amount of information they can use to perform work is the quantum deficit.
We have seen that in the process of diluting e-bits inevitably some information corresponding to the entropy $S(\rho_{ab})$ is lost to the environment.
In our approach, the environment is represented by $c$.
 Therefore the loss of entanglement is, \textit{de facto}, associated to part of this information lost to the environment and is quantified by $\Delta_{a|c}$.

To summarize our results provide strong evidences that irreversibility must happen for all mixed, not PP, entangled states. We have shown that such a counter-intuitive possibility would necessarily imply other very counter-intuitive properties.
For instance one possibility is having $\delta_{a|c}>0$ and $\Delta_{a|c}=0$.
In this case the non-additivity of EOF would be responsible for irreversibility.
Other possibility is that, to obtain $E^{\cal D}$, it is necessary to optimize the coherent information over an arbitrary large number of copies of the entangled state.
Moreover we have shown irreversibility for the important family of 1-MC states.
In addition we calculate $E^{\cal D}$, quantum discord and the relative entropy of entanglement for them and, further, we have shown that there is no bound entangled and that $S_{a|b}=0$ is a necessary and sufficient separability criteria for this family.

The Authors acknowledge support from FAPESP and CNPq through INCT-IQ.

\section*{Supplementary Material: Proofs of Theorems}

Here we show the proofs of the Theorems and Lemma of the main paper. We restate them for clarity.

\textit{Lemma 1}: For every mixed entangled state $\rho_{ab}$, such that $\rho_{ab}$ is not a PP state, we have
\begin{equation}
E^{\cal F}(\rho_{ab})>I_{C}(\rho_{ab}).
\label{Theo1a}
\end{equation}
Furthermore, if $E^{\cal F}$ is additive for $\rho_{ab}$, then
\begin{equation}
E^{\cal C}(\rho_{ab})>I_{C}(\rho_{ab}).
\label{Theo1b}
\end{equation}

\textit{Proof}: We first show that if $\delta_{a|c}(\rho_{ac})=0$ (or $\delta_{b|c}(\rho_{bc})=0$) then $\rho_{ab}$ is PP. Then the result will follows easily from Eq. (7) of the main article.

 It is known from Refs. \cite{zurek01,horodeckietal05-deficit-pra,vedral10-prl} that 
$\delta_{a|c}(\rho_{ac})=0$ if and only if one can write $\rho_{ac}$ as
\[
\rho_{ac}=\sum_{i}p_{i}\rho_{i}^{(a)}\otimes|i^{(c)}\rangle \langle i^{(c)}|,
\]
where $\rho_{i}^{(a)}$ is an arbitrary state of $a$, the set $\{|i^{(c)}\rangle\}$ is an orthonormal basis for $c$ and
$p_{i}$ is the probability of $i$ being measured in $c$.
Now let $|\lambda_{ij}\rangle$
be the $j$-th eigenvector of $\rho_{i}^{a}$. As  $\{|\lambda_{ij}\rangle|i\rangle\}$
span the eigenvectors of $\rho_{ac}$, it can be purified using the Schmidt decomposition to
\[
|\psi_{abc}\rangle=\sum_{ij}\sqrt{p_{i}\lambda_{ij}}|\lambda_{ij}\rangle|b_{ij}\rangle\left|i\right\rangle.
\]
where $\{|b_{ij}\rangle\}$ are states of the quantum subsystem belonging to Bob.
Thought in general $\{|\lambda_{ij}\rangle\}$ may not be orthogonal for different $i$ (the support of the states $\rho_i^{(a)}$ are arbitrary), the states $|b_{ij}\rangle$ are, since they are the eigenvectors of a Schmidt decomposition. So
 Bob is able to distinguish the $i$ components of the state and to use them
as the flags of a PP state. Thus we can write
\[
\rho_{ab}=\sum_{i}p_{i}|\psi_{i}\rangle\langle\psi_{i}|\otimes|f_{i}\rangle\langle f_{i}|,
\]
 where $|\psi_{i}\rangle=\sum_{j}\sqrt{\lambda_{j}}|\lambda_{ij}\rangle|b_{ij}\rangle$ and one can check that $\rho_i^{(a)}$ are the reduced states of $|\psi_i\rangle$.
Now Eq. (\ref{Theo1a}) follows from Eq. (7) of the main article,
\[
E^{\cal F}(\rho_{ab}) = \delta_{a|c}(\rho_{ac})-S_{a|b}(\rho_{ab})
\]
Further, when EOF is additive for the specific state $\rho_{ab}$, we have $E^{\cal F}(\rho_{ab})=E^{\cal C}(\rho_{ab})$ and Eq. (\ref{Theo1b}) follows. $\hfill\blacksquare$

\textit{Theorem 1}: 
Let $\rho_{ab}$ an entangled mixed state, if 
\begin{eqnarray}
E^{\cal C}(\rho_{ab})=\frac{1}{n}E^{\cal F}(\rho_{ab}^{\otimes n})
\label{ap:CondA}
\\
E^{\cal D}(\rho_{ab})=\max_{V}\frac{1}{k}I_{C}\left(V\rho^{\otimes k}\right)
\label{ap:CondB}
\end{eqnarray}
for some finite $n$ and $k$, then entanglement is irreversible for $\rho_{ab}$, i. e., $E^{\cal C}(\rho_{ab})>E^{\cal D}(\rho_{ab})$.


\textit{Proof}: The proof follows by contradiction. Let us make the following three hypothesis:

(H1) EOF is additive for some finite number of copies of $\rho_{ab}$. That is, there exists some $n$ such that, 
\[
E^{\cal C}(\rho_{ab})=\frac{1}{n}E^{\cal F}(\rho_{ab}^{\otimes n}).
\] 

(H2) It is possible to attain $E^{\cal D}(\rho_{ab})$ with the LOCC $V$ action on a finite number of copies $k$ in Eq. (4) of the main paper. That is, for some finite $k$,
\[
E^{\cal D}(\rho_{ab})=\frac{1}{k}\max_V I_{C}(V \rho_{ab}^{\otimes k}),
\]
where $V$ is some LOCC in the space of $k$ copies of $\rho_{ab}$.

(H3) Entanglement is reversible for $\rho_{ab}$. That is, 
\[
E^{\cal C}(\rho_{ab})=E^{\cal D}(\rho_{ab}).
\]

We will show that these three hypothesis are not compatible with inequality (\ref{Theo1b}) that we have just proved in Lemma 1. Let $\rho_{ab}$ a state satisfying the three hypothesis and let $V^\prime$ being a LOCC given the maximum in (H2), that is,
\[
E^{\cal D}(\rho_{ab})=\frac{1}{k} I_{C}(V^\prime \rho_{ab}^{\otimes k})
\]
Once that, by (H2),  $V^\prime$ cannot decrease the average entanglement cost of $\rho_{ab}$, since in this case it would also decrease $E^{\cal D}$, and it cannot increase $E^{\cal C}$ since it is a LOCC, we must have
\[
E^{\cal C}(V^\prime \rho_{ab}^{\otimes k})=E^{\cal C}(\rho_{ab}^{\otimes k}).
\]
Since $E^{\cal F}$ also cannot increase under LOCC, we  must have,
\[
E^{\cal F}\left[(V^\prime \rho_{ab}^{\otimes k})^{\otimes n}\right]
\le E^{\cal F}\left[(\rho_{ab}^{\otimes k})^{\otimes n}\right]
 = E^{\cal C}\left[(\rho_{ab}^{\otimes k})^{\otimes n}\right], 
\]
where the last equality comes from (H1). On the other hand, if $E^{\cal F}[(V^\prime \rho_{ab}^{\otimes k})^{\otimes n}]$ was strictly smaller than $E^{\cal C}(\rho_{ab}^{\otimes nk})$, we would have
\begin{eqnarray*}
E^{\cal C}(\rho_{ab}^{\otimes nk}) & > &  E^{\cal F}\left[(V^\prime \rho_{ab}^{\otimes k})^{\otimes n}\right] \\
& \ge & E^{\cal C}\left[(V^\prime \rho_{ab}^{\otimes k})^{\otimes n}\right]\\
& \ge & E^{\cal D}\left[(V^\prime \rho_{ab}^{\otimes k})^{\otimes n}\right]\\
& = & E^{\cal D}(\rho_{ab}^{\otimes nk}).
\end{eqnarray*}
But then we would have $E^{\cal C}(\rho_{ab})>E^{\cal D}(\rho_{ab})$ and this contradicts (H3). So all the measures in the previous expression must be equal, in particular 
\begin{equation}
E^{\cal C}(V^\prime \rho_{ab}^{\otimes k})=\frac{1}{n} E^{\cal F}\left[(V^\prime \rho_{ab}^{\otimes k})^{\otimes n}\right]
=E^{\cal C}(\rho_{ab}^{\otimes k}).
\label{EcNotChangedByV}
\end{equation}
In other words, if EOF is additive for $\rho_{ab}^{\otimes n}$, it must be  also additive for $(V^\prime \rho_{ab}^{\otimes k})^{\otimes n}$. Now, by Eq. (\ref{Theo1b}), we have
\[
E^{\cal C}\left[(V^\prime \rho_{ab}^{\otimes k})^{\otimes n}\right] > I^C\left[(V^\prime \rho_{ab}^{\otimes k})^{\otimes n}\right]
\]
Then, by Eq. (\ref{EcNotChangedByV}) and by (H2),
\[
E^{\cal C}(\rho_{ab}) > E^{\cal D}(\rho_{ab}) 
\]
But this inequality contradicts the hypothesis  (H3). So all the three hypothesis, (H1), (H2) and (H3), can not be true simultaneously for any mixed entangled state $\rho_{ab}$ that is not a PP state. $\hfill\blacksquare$

In order to proceed to the proof of the Theorem 2, we need to introduce two important
 measures of entanglement: the distillable secret key $K$ and the relative entropy of entanglement
$R$ \cite{VedralPlenio98-RelativeEntropy-pra}. The first is the
rate of conversion of many copies of a state $\rho_{ab}$ into bits of
secret correlation between Alice and Bob  \cite{HorodeckiReview09}. 
In fact $K \ge E^{\cal D}$, since one can always obtain one bit of secret correlation from one e-bit. Surprisingly,
it is possible to have $K>0$ for entangled states where $E^{\cal D}=0$
\cite{horodecki05-keyfromboundent-prl}.
In general, however, it is believed that there are entangled states from which no secret key can be extracted.

The relative entropy of entanglement and its regularization are defined as
\[
{R}(\rho)=\min_{\sigma\in \mathtt{Sep}} S(\rho\parallel\sigma)
\]
and
\[
R^{\infty}=\lim_{n\rightarrow\infty}\frac{R(\rho^{\otimes n})}{n},
\]
where $\sigma$ belongs to the set of separable states and ${S(\rho\parallel\sigma)}$ is the usual
relative entropy.
Furthermore, $R$ and $R^{\infty}$ also provide useful bounds for $E^{\cal C}$ and $E^{\cal D}$ \cite{VedralPlenio98-RelativeEntropy-pra,horodecki05-keyfromboundent-prl,HorodeckiReview09}, 
\begin{equation}
E^{\cal F}\geq E^{\cal C}\ge R^{\infty}\ge K\geq E^{\cal D}\ge I_C
\label{inequalities}
\end{equation}
and
\begin{equation}
E^{\cal F}\ge R\geq R^{\infty}.
\label{inequalities2}
\end{equation}

We can now restate the Theorem 2 and prove it. The statement is more general than the one given in the Letter and involve the expression for the measures $K$ and $R$.

\textit{Theorem 2}: For every mixed state $\rho_{ab}$,  and its
respective purification $|\psi_{abc}\rangle$ such that the complementary
state $\rho_{ac}$ is separable, the entanglement between $a$ and $b$ is irreversible,
\[
E_{ab}^{\cal C} > E^{\cal D}_{ab},
\]
\begin{equation}
E^{\cal D}_{ab} = \delta_{a|b} = \Delta_{a|b} = R_{ab} = R_{ab}^\infty = K_{ab} = -S_{a|b}.
\label{Th2bAp}
\end{equation}
and
\begin{equation}
\Delta_{a|c}(\rho_{ac})=E^{\cal C}(\rho_{ab})-E^\mathcal{D}(\rho_{ab}),
\label{eq:EcDefEdAp}
\end{equation}

 \textit{Proof}: First we prove Eq. (\ref{Th2bAp}). When $\rho_{ac}$ is separable, we have $E^{\cal F}_{ac}=E^{\cal C}_{ac}=0$.
So exchanging subsystems $b$ and $c$ in Eqs. (7) and (8) of the main paper we have
\[
\delta_{a|b}=-S_{a|b}
\]
and
\[
\Delta_{a|b}= -S_{a|b}.
\]
However we know from Refs. \cite{horodeckietal05-deficit-pra,vedral10-prl} that
$\delta_{a|b}$ is an upper bound to $R$ which, on its turn, is an upper bound to $R^\infty_{ab}$, $K_{ab}$, $E^{\cal D}_{ab}$ and $-S_{a|b}$ according to Eq.  (\ref{inequalities}) and Eq. (\ref{inequalities2}).
Therefore all these quantities must be equal to $-S_{a|b}$ and Eq. (\ref{Th2bAp}) follows.

Now it is easy to prove that $E_{ab}^{\cal C} > E^{\cal D}_{ab}$. By Lemma 1, we have that
\[
E^{\cal F}_{ab}>-S_{a|b}=E^{\cal D}_{ab}.
\]
From Ref. \cite{DevetakWinter04-ccr}, we know that $J_{a|c}$ is additive for separable states.
Then, from Ref. \cite{koashiwinter04-physreva.69.022309}, we have
\begin{equation}
E_{ab}+J_{a|c}=S_a.
\label{KoashiWinter}
\end{equation}
Once that $S_a$ is additive, it implies that $E_{ab}^{\cal F}$ is additive for $\rho_{ab}$ when $J_{a|c}$ is additive for the complementary state $\rho_{ac}$.
 So  $E_{ab}^{\cal F}=E_{ab}^{\cal C}$ and $E_{ab}^C>{\cal D}_{ab}$.
$\hfill\blacksquare$

Notice that, once $\delta_{a|b}=-S_{a|b}$, it also implies that $S_{a|b}$ is always negative for the family of states $\rho_{ab}$ such that the complementary state $\rho_{ac}$ is separable.
Therefore $\rho_{ab}$ is separable \textit{if and only if} $\delta_{a|b}=-S_{a|b}=0$. So $S_{a|b}=0$ provides a necessary and sufficient separability criteria for 1-MC. In addition, Theorem 2 also proves that the relative entropy of entanglement is additive for this family of states and that there is no bound entangled states in.

Now we proceed to the prove of Corollary 1.

\textit{Corollary 1}: A type $B$ reversible mixed state $\rho_{ab}$ exists \textit{if and only if} there exists a bound entangled state $\rho_{ac}$, with $K_{ac}=0$ such that $\delta_{a|c}>0$ and $\Delta_{a|c}=0$.

\textit{Proof:} Suppose the state is of type $B$, i. e. it satisfies the condition (\ref{ap:CondB}) in Theorem 1.
Then there is some finite $k$ and some optimal distillable state $\sigma_{ab}$ such that ${\sigma_{ab}=V\rho_{ab}^k}$. If $\rho_{ab}$ is reversible, then $\sigma_{ab}$ is necessarily reversible too. So we can suppose, without loss of generality, that $\rho_{ab}$ is already an optimal distillable state satisfying Eq. (\ref{eq:EcDefEdAp}). Now we have three possibilities for the complementary states $\rho_{ac}$:

(i) It is separable, but them Theorem 2 shows that it is irreversible;

(ii) It is entangled and have some distillable secret key. In this case we have
\[
\Delta_{a|c}\geq R^\infty_{ac} \geq K_{ac} > 0.
\]
As $\rho_{ab}$ satisfies Eq. (\ref{eq:EcDefEdAp}) and $\Delta_{a|c}$ is strictly greater than zero, $\rho_{ab}$ is also irreversible.

(iii) It is a bound entangled state with $K_{ac}=0$. That is the condition stated in the corollary.

Conversely, suppose that there is a $\rho_{ac}$ such that $\Delta_{a|c}=0$, then it is a bound entangled state, $K_{ac}=0$ and, by Eq. (\ref{eq:EcDefEdAp}),
\[
E^{\cal C}=-S_{a|b}=E^{\cal D}.
\]
That is entanglement is reversible.
$\hfill\blacksquare$

Remark that being a bit more technical we have a stronger result than the Corollary 1 presented in the Letter. Since there are states with $K>0$ but $E^{\cal D}=0$ \cite{horodecki05-keyfromboundent-prl} and $K$ is also a lower bound for $R$, we can replace $E^{\cal D}$ for $K$ in the statement of the Corollary. 

We can also prove more one simple corollary that was not presented in the main paper.

\textit{Corollary 2}: Let $\rho_{ab}$ be a mixed not PP entangled state
and $|\psi_{abc}\rangle$ its purification. Then, if $K(\rho_{ac})>0$
and $K(\rho_{bc})>0$, then
\[
E^{C}(\rho_{ab})>I_{C}(\rho_{ab}).
\]

Proof. If $K(\rho_{ac})>0$, then we know that ${R^{\infty}(\rho_{ac})>0}$ since $R^{\infty}\ge K$. But $\Delta_{a|c}\ge R^{\infty},$ therefore
${\Delta_{a|c}(\rho_{ac})>0}$ and by Eq. (8) of the main article
we have ${E^{C}(\rho_{ab})>-S_{a|b}}$ $(\rho_{ab})$.
Similar argument holds for $\rho_{bc}$ and $E^{C}(\rho_{ab})>-S_{b|a}(\rho_{ab})$.
$\hfill\blacksquare$

\bibliographystyle{apsrev4-1} 
\bibliography{EntIrr-abr}

\end{document}